\documentclass[aps,prb,showpacs,preprint,floatfix]{revtex4}

\usepackage{graphicx}
\usepackage{dcolumn}
\usepackage{amsmath}
\usepackage[latin1]{inputenc}

\begin{document}

\title{Direct Correlation between $1/f$-magneto-noise and magnetoresistance in La$_{0.7}$Sr$_{0.3}$MnO$_3$
and (La$_{0.5}$Pr$_{0.2}$)Ba$_{0.3}$MnO$_3$ manganites}

\author{D. S. Rana}
\email{dhanvir_rana1@rediffmail.com}
\altaffiliation{Present address:
Institute of Laser Engineering,
Osaka University,
2-6 Yamadaoka, Suita,
Osaka 565-0871, Japan}
\affiliation{
Division of Superconductivity and Magnetism,
University of Leipzig,
04103 Leipzig, Germany\\ and\\
Tata Institute of Fundamental Research,
Homi Bhabha Road, Colaba, Mumbai - 400 005, India}
\author{M. Ziese}
\affiliation{
Division of Superconductivity and Magnetism,
University of Leipzig,
04103 Leipzig, Germany}
\author{S. K. Malik}
\affiliation{
Tata Institute of Fundamental Research,
Homi Bhabha Road, Colaba, Mumbai - 400 005, India}

\date{\today}

\begin{abstract}
Temperature- and magnetic field-dependent electrical noise and electrical resistivity
measurements were carried out on epitaxial thin films of a large bandwidth
La$_{0.7}$Sr$_{0.3}$MnO$_3$ and a disordered intermediate bandwidth (La$_{0.5}$Pr$_{0.2}$)Ba$_{0.3}$MnO$_3$
manganite system. The power spectral density was dominated by $1/f$-noise. This $1/f$-noise was observed
to follow the overall temperature dependence of the resistivity.
Moreover, in these compounds the magneto-noise effect was found to be of nearly the same magnitude as
the magnetoresistance. This direct correlation between magneto-noise and magnetoresistance suggests that
the enhanced $1/f$-noise has its origin in intrinsic charge-carrier density fluctuations.
\end{abstract}
\maketitle
Noise measurements are an important tool to understand various underlying transport mechanisms in
different classes of metallic and semiconducting materials. Fluctuations in the charge transport
of condensed matter systems arise from various relaxation processes of the
charge carriers, defects, groups of defects, etc., see Refs.~[\onlinecite{dutta1981,weissman1998}] for an overview.
The noise power spectral density $S_V$ in general shows a $1/f^\alpha$-dependence, where $f$
is the frequency of the fluctuation and $\alpha$ is an exponent close to unity, and is referred to
as $1/f$-noise.\cite{dutta1981,weissman1998} Amongst various oxides, mixed-valent manganites are a class of modern era
technological materials, which are rich in exhibiting exotic electronic and magnetic phases in the
temperature-magnetic field phase diagram.\cite{colossal,coey1999} Previous work has established that
$1/f$-electrical noise in mixed-valent manganites showing colossal magnetoresistance is maximal
in the vicinity of the metal-insulator transition temperature.\cite{anane2000,rajeswari1996,rajeswari1998,ahlers1996,raquet1999,reutler2000}
In addition, phase separation in manganites was indicated by measurements of random telegraph noise, where
a continuous switching between insulating and metallic phases near the metal-insulator transition temperature
shows the dynamic nature of phase separation processes in these compounds.\cite{raquet2000,bid2003}

Near the Curie temperature phases with different conductivities and/or carrier concentrations exist in manganites. Recently,
vibronic phase segregation into hole-rich, itinerant and hole-poor, polaronic regions was reported in orthorhombic
La$_{2/3}$(Ca$_{1-x}$Sr$_x$)$_{1/3}$MnO$_3$. \cite{rivadulla2006} This type of phase separation is driven by the coupling
of the charge carriers to Jahn-Teller distortions. Another kind of charge-carrier fluctuations was discussed
by Alexandrov and Bratkovsky. \cite{alexandrov1999,alexandrov2006} In this case the charge carriers are supposed
to be polarons of oxygen $p$-character which are bound into heavy immobile bi-polarons near the transition temperature.
This mechanism leads to a strong temperature and field dependent polaron density that determines the behaviour of
the resistivity.\cite{alexandrov1999,zhao2000} However, optical spectroscopy studies failed to find evidence
for bi-polaron formation, whereas small polaron formation in La$_{0.7}$Ca$_{0.3}$MnO$_3$ and large polaron
formation in La$_{0.7}$Sr$_{0.3}$MnO$_3$ was observed. \cite{hartinger2006} Furthermore, the significance of quenched disorder has been pointed out
\cite{burgy2004} and the existence of a Griffith's phase has been reported.\cite{salamon2002,deisenhofer2005}
Quenched disorder can also be incorporated in the bi-polaron model leading to phase segregation
into itinerant polaron and immobile bi-polaron phases near the Curie temperature. \cite{alexandrov2006} At constant temperature
the application of a magnetic field drives the electronic system towards an itinerant state with an increasing mobile
carrier density. The susceptibility of the carrier concentration to magnetic fields is highest near the Curie temperature.
Experimentally, a carrier density change as a function of magnetic field near the transition temperature is difficult to detect
with conventional Hall-effect measurements, since high fields are necessary to saturate the anomalous Hall effect.
\cite{westerburg2000} Here an alternative method for the study of charge-density variations is proposed.

In spite of the fact that several studies on electrical noise in manganites found evidence for a correlation
between the $1/f$-noise and the magnitude of the resistivity as a function of temperature, the influence of
the magnetic field on the $1/f$-noise and its correlation with the magnetoresistance has not been
clearly demonstrated experimentally.\cite{anane2000,rajeswari1996,rajeswari1998,ahlers1996,raquet1999,reutler2000}
Magnetic field dependent studies, however, would be suitable to further elucidate the origin of the $1/f$-noise
in the manganites, since it would be possible to further characterize the crossover from localized
to itinerant charge carriers. In this paper we report magnetic field- and temperature-dependent $1/f$-electrical noise and
resistivity studies on two distinct manganites, namely, an intermediate $e_g$-electron bandwidth A-site
disordered manganite of composition (La$_{0.5}$Pr$_{0.2}$)Ba$_{0.3}$MnO$_3$ (LPBMO) and a standard large $e_g$-electron
bandwidth manganite, La$_{0.7}$Sr$_{0.3}$MnO$_3$ (LSMO).
These samples have peak resistivity temperatures of $T_p \simeq 200$~K in case of LPBMO and $T_p \simeq 375$~K
in case of LSMO. In spite of the large difference in the peak temperatures of the
two systems, we show that the magneto-noise and magnetoresistance are of nearly the same magnitude and
that the $1/f$-noise follows nearly the same trend as the electrical resistance. This indicates that the
$1/f$-noise and the electrical resistance may have the same physical origin.

Epitaxial thin films of (La$_{0.5}$Pr$_{0.2}$)Ba$_{0.3}$MnO$_3$ (LPBMO) and La$_{0.7}$Sr$_{0.3}$MnO$_3$ (LSMO)
were deposited on $(001)$ oriented LaAlO$_3$ single-crystal substrates using Pulsed Laser
Deposition (PLD) with a KrF excimer laser. The details of the synthesis and other transport properties
of the LPBMO film (thickness $\simeq 200$~nm) were reported elsewhere.\cite{rana2005,mavani2004}
The film investigated here was grown at optimized deposition parameters as described in [\onlinecite{mavani2004}],
i.e.~a fluence of $\sim 3.1$~J/cm$^2$,
a growth rate of $0.33$~nm/s, a substrate temperature of $830^\circ$C, an O$_2$ partial pressure of 400~mTorr and
a substrate to rotating target distance of $4.2$~cm.
The LSMO film (thickness $\simeq 150$~nm) was deposited at a substrate temperature of 700$^\circ$C in an oxygen
partial pressure of 0.13~mbar.\cite{ziese2002} The LPBMO and LSMO thin films were found to be $(101)$ and $(001)$
oriented, respectively. Electrical resistivity, as a function of temperature and magnetic field, was measured
in a dc four probe configuration using a Physical Property Measurement System (PPMS, Quantum Design).
An ac five probe method \cite{scofield1987} was used for the noise measurement. A voltage signal with a frequency
of 679~Hz was fed to the sample from the internal oscillator of a lock-in amplifier (Stanford Research Systems, SRS,
Model SR830). The compensated signal from the sample was demodulated and amplified by the same lock-in amplifier
using a time constant of 1~ms and a band-pass filter setting of 24~db. The analog signals of both the in-phase
and out-of-phase component were fed into a dual channel spectrum analyzer (SRS, Model SR785) and the power spectral
density was measured in the frequency range between 100~mHz and 25~Hz. The in-phase noise was always considerably larger
than the out-of-phase (background) noise. The measurements were performed in a He flow-cryostat
equipped with a superconducting solenoid up to 9~T. Measurements of the voltage dependence of the power spectral density
were performed and a quadratic dependence of the $1/f$-noise on bias voltage was found. This shows that the
recorded noise power spectral density is intrinsic to the samples.\cite{raquet1999}

\begin{figure}
\begin{center}
\vspace*{-0.0cm}
\includegraphics[width=0.75\textwidth]{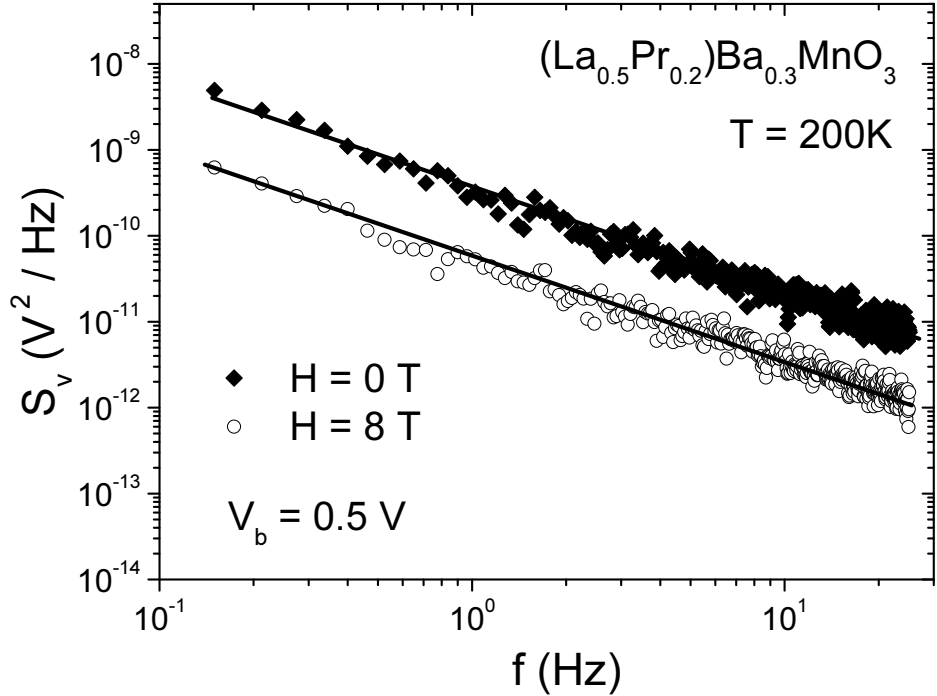}
\vspace*{-0.0cm}
\end{center}
\caption{Frequency $f$ dependence of the noise power spectral density $S_V$ for the (La$_{0.5}$Pr$_{0.2}$)Ba$_{0.3}$MnO$_3$
film in magnetic fields $\mu_0H$ of 0 and 8~T at a temperature of 200~K. The bias voltage was 0.5~V.}
\label{fig1}
\end{figure}
\begin{figure}
\begin{center}
\vspace*{-0.0cm}
\includegraphics[width=0.75\textwidth]{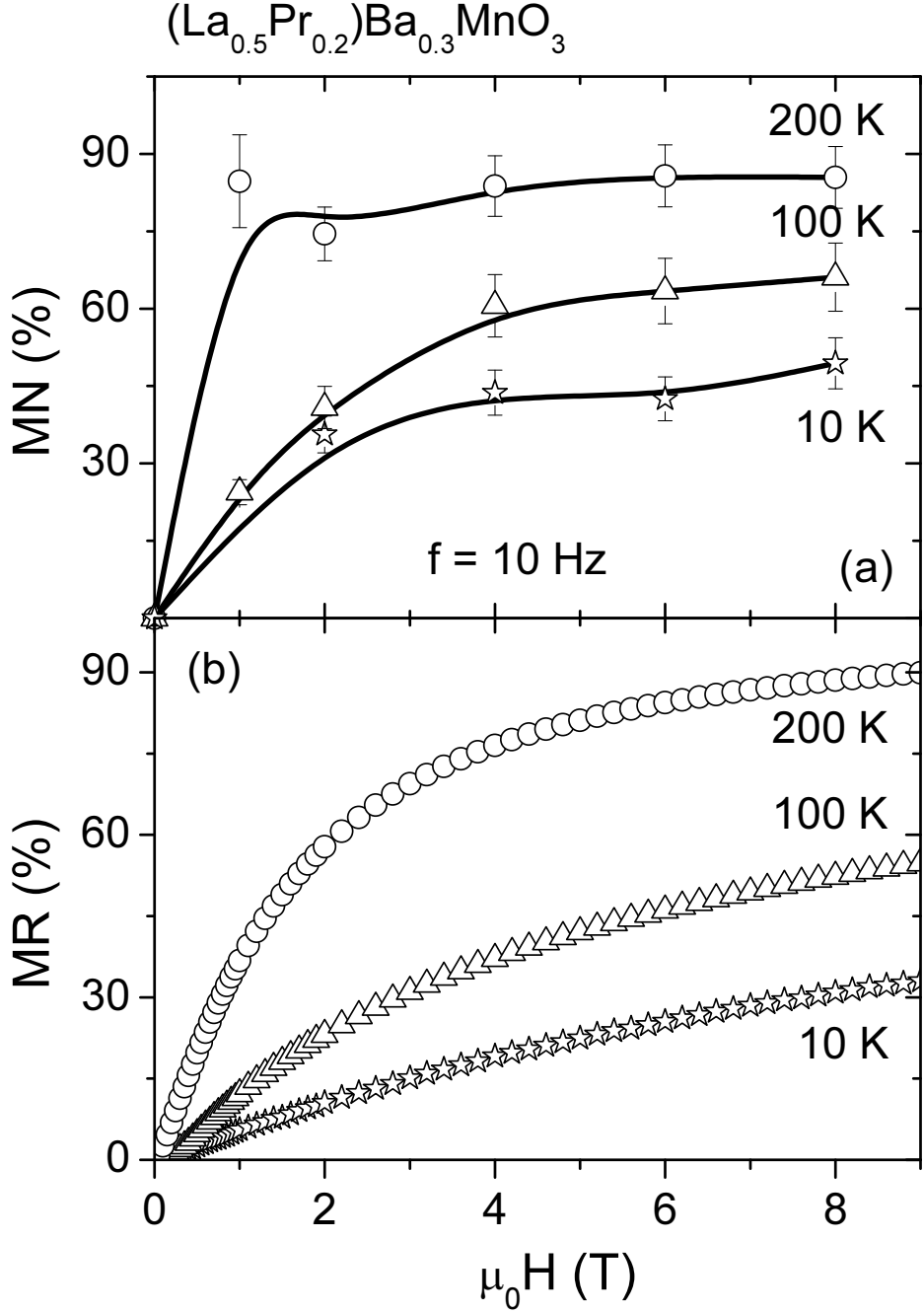}
\vspace*{-0.0cm}
\end{center}
\caption{(a) Magneto-noise $MN$ at a frequency of 10~Hz and (b) magnetoresistance
as a function of magnetic field for the (La$_{0.5}$Pr$_{0.2}$)Ba$_{0.3}$MnO$_3$ thin film at 200~K, 100~K and 10~K.
The solid lines in (a) are guides to the eye.}
\label{fig2}
\end{figure}
\begin{figure}
\begin{center}
\vspace*{0.0cm}
\includegraphics[width=0.75\textwidth]{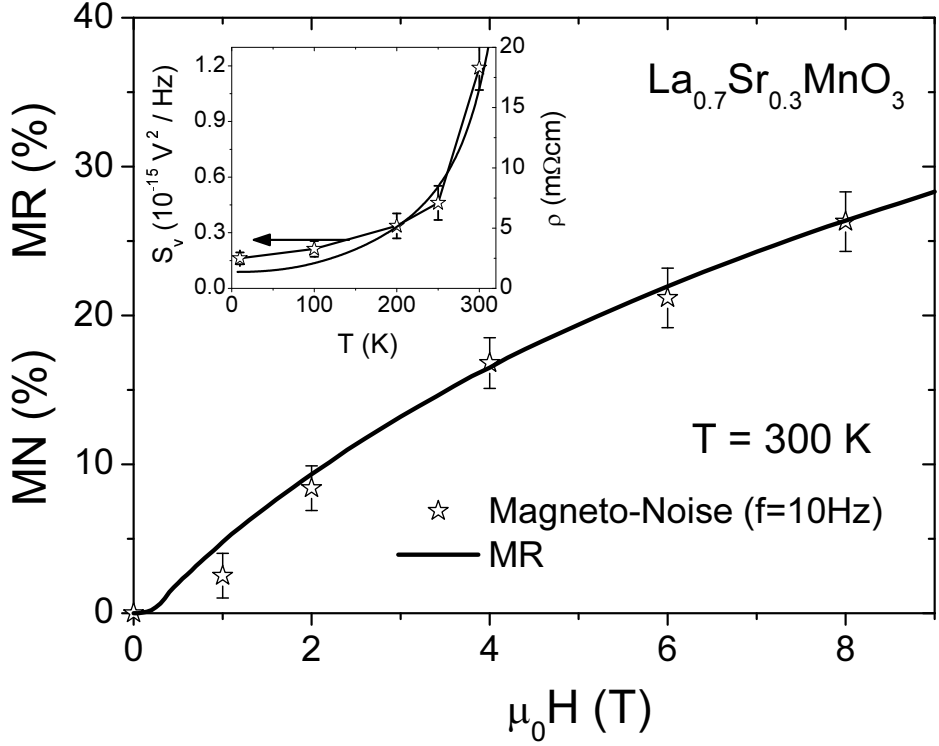}
\vspace*{0.0cm}
\end{center}
\caption{Magneto-noise $MN$ at a frequency of 10~Hz and magnetoresistance as a function of magnetic field for the
La$_{0.7}$Sr$_{0.3}$MnO$_3$ thin film at 300~K. The left and the right scale of the inset
show the temperature dependence of the noise power spectral density $S_V$ and the electrical resistivity $\rho$, respectively.}
\label{fig3}
\end{figure}

Figure~\ref{fig1} shows a typical frequency dependence of the power spectral density $S_V$ in zero magnetic
field and in a field of 8~T for the LPBMO film at a temperature of 200~K, i.e. in close vicinity
of its peak resistivity temperature. Fits of a power law to the power spectral density show a $S_V \propto 1/f^\alpha$
dependence with an exponent $\alpha \simeq 1$. Additionally, it is noteworthy that there is a large suppression
of $S_V$ under the application of a magnetic field. This reveals a large sensitivity of the $1/f$-noise to the
magnetic field near $T_p$. In order to correlate the magnetic field
dependence of the $1/f$-noise and the electrical resistance, the magneto-noise
\begin{equation}
MN = \frac{S_V(0)-S_V(H)}{S_V(0)}
\label{mn}
\end{equation}
and the magnetoresistance
\begin{equation}
MR = \frac{R(0)-R(H)}{R(0)}
\label{mr}
\end{equation}
are shown in Fig.~\ref{fig2} as a function of magnetic field at various temperatures for the intermediate
bandwidth LPBMO film. It is seen that the magneto-noise is nearly of the same magnitude as the magnetoresistance
at all the temperatures investigated and the overall behaviour of the magneto-noise (shown at 10~Hz) and
the magnetoresistance are similar. Magneto-noise curves at other frequencies show the same field dependence, since
the exponent $\alpha$ is independent of field. The magnetoresistance and magneto-noise for the large
bandwidth LSMO film are shown in Fig.~\ref{fig3} at a temperature of 300~K. The agreement between the field-dependent
$MN$- and $MR$-curves for the two effects are striking. These observations point towards a common origin of
these two properties. This is a novel feature of the transport properties of the manganites which has not yet been
elucidated in all the previous magnetic field dependent noise studies.\cite{rajeswari1998,ahlers1996,raquet1999,reutler2000}
For instance, the magnetic field dependent $1/f$-noise studies on La$_{0.7}$Sr$_{0.3}$MnO$_3$ by Raquet {\it et al.}\cite{raquet1999}
show a magneto-noise effect of $\sim 60$\% versus a magnetoresistance of $\sim15$\% in
a field of 5~T at 300~K; in La$_{0.7}$Ca$_{0.3}$MnO$_3$ films a suppression
of the noise peak near the metal-insulator transition followed by a saturation of the magneto-noise
in fields as low as 2~T was observed.\cite{reutler2000}
\begin{figure}
\begin{center}
\vspace*{-0.0cm}
\includegraphics[width=0.75\textwidth]{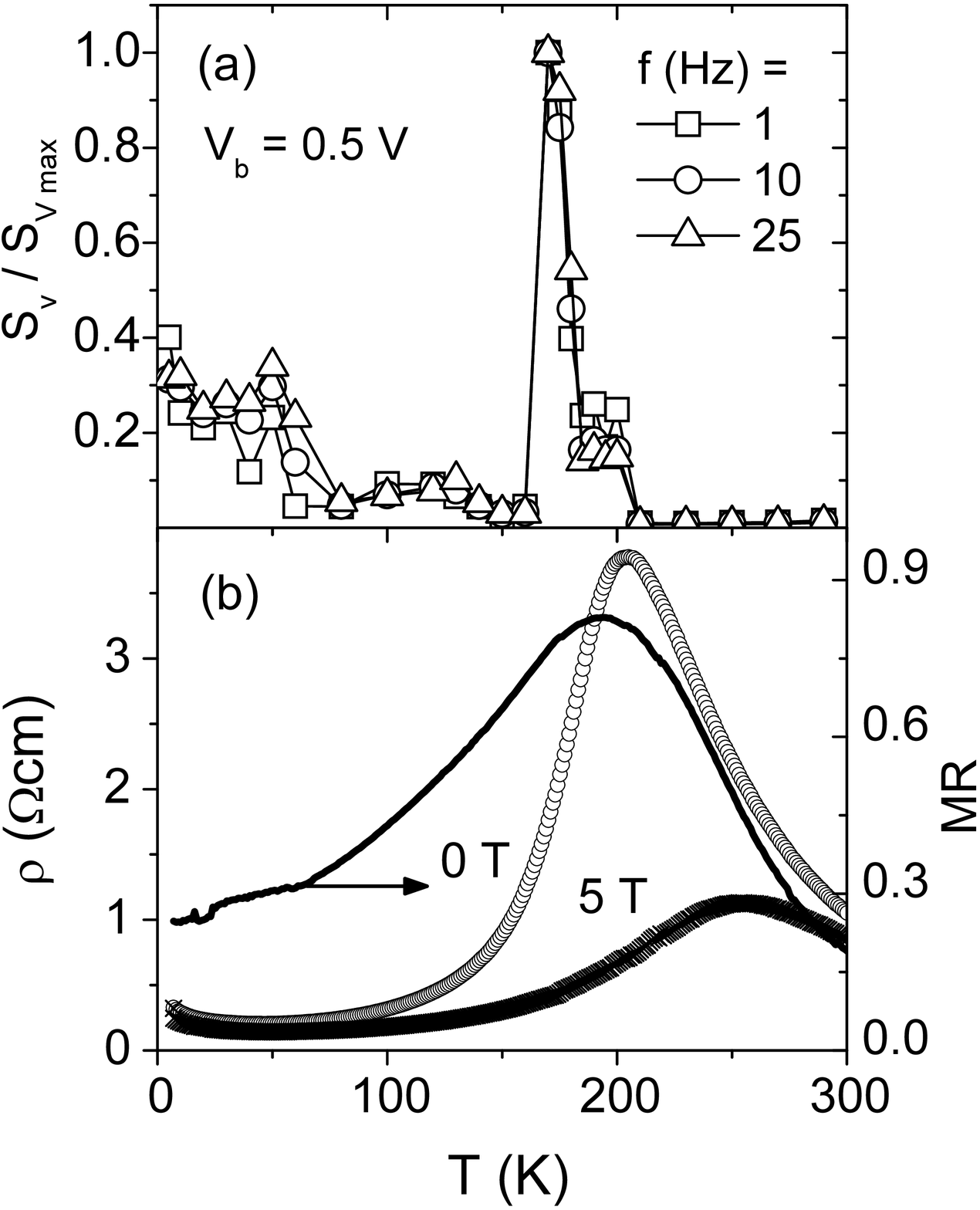}
\vspace*{-0.0cm}
\end{center}
\caption{(a) Noise power spectral density $S_V$ normalized to its maximum values versus temperature at frequencies of
1~Hz, 10~Hz and 25~Hz for the (La$_{0.5}$Pr$_{0.2}$)Ba$_{0.3}$MnO$_3$ film.
(b) Temperature dependence of the electrical resistance $R$ and magnetoresistance $MR$
in a field of 5~T.}
\label{fig4}
\end{figure}
In order to obtain more information on the origin of the direct correlation of the field-dependent
sensitivity of the $1/f$-noise and the field-dependence of the electrical resistance,
a detailed correlation of the temperature-dependent $1/f$-electrical noise in zero field with the
zero field resistance of the LPBMO sample has been experimentally established from the data presented
here. The noise is lowest in the semiconducting and paramagnetic region above $T_C$, whereas it
peaks in the vicinity of the metal-insulator transition. More precisely, the noise is found to be highest
at a temperature where the magnetoresistance and the first derivative of the
zero field resistivity are maximum, just below the peak resistivity temperature.
The noise is higher in the metallic region as compared to that in the
region above $T_p$. Further, in the metallic region, the noise continues to rise as the temperature decreases,
which is similar to the low temperature rise in resistivity, thus indicating a simultaneous low temperature
evolution of additional sources of noise and resistivity. The noise power spectral density $S_V$ at 1~Hz, 10~Hz
and 25~Hz, the resistance and MR are plotted as a function of temperature in Fig.~\ref{fig4}. The scaled power spectral
densities $S_V/S_{Vmax}$ agree rather well and indicate the scatter in the data. Similar behaviour is observed in
the LSMO film, see inset to Fig.~\ref{fig3}. Throughout the temperature range investigated, for both LSMO and
LPBMO samples, the noise power spectral density as a function of frequency obeys a power law
$S_V \propto 1/f^\alpha$ with $0.9 < \alpha < 1.25$ both in the absence and presence of an applied magnetic field.

Despite the fact that the magnitude of noise in LPBMO is rather large as compared to that in LSMO, both of
these samples exhibit qualitatively similar relations between the $1/f$-noise and the electrical resistivity.
In the absence of any detailed theory of $1/f$-noise in manganites, here very general expressions to quantify
the parameter dependence of the noise and the resistivity will be used. The power spectral density is given by
Hooge's law\cite{hooge1969}
\begin{equation}
S_V = \gamma\frac{V^2}{n\Omega} \frac{1}{f^\alpha}
\label{hooge}
\end{equation}
as a function of the voltage $V$ applied across the sample, the number $n$ of carriers, the sample volume $\Omega$
and the Hooge parameter $\gamma$ that characterizes the strength of noise sources. For the resistivity the Drude
expression is used, since this represents a general principle, namely the factorization into a carrier density and a mobility:
\begin{equation}
\rho = \frac{m^*}{n{\rm e}^2\tau}
\label{drude}
\end{equation}
with the effective mass $m^*$, electronic charge e and relaxation time $\tau$.

At first the zero field values are discussed and estimates for the Hooge parameter $\gamma$ are derived
to facilitate comparison to previous work on $1/f$-noise in manganites. Within a factor of 3-5 Hall effect
measurements \cite{jakob1998,ziese1999,westerburg2000}
indicate a carrier concentration $n \sim 5\times10^{21}$~cm$^{-3}$ for an optimally hole-doped system. Since
this value is inferred from Hall measurements in a finite field, the zero field carrier concentration might actually be smaller
than this value. However, in order to obtain an order of magnitude estimate, this carrier concentration is used in the following.
A value of $\gamma \simeq 1-2\times 10^6$
is estimated at 200~K for the LPBMO film and $\gamma \simeq 700-1000$ at 300~K for the LSMO film. Both
values are much larger than typical values\cite{dutta1981,weissman1998} for metallic systems of $\gamma \simeq 10^{-3}$.
Although $\gamma$ for LPBMO is very large compared to that observed in metals and semiconductors,
it is of the same order as that reported for Pr$_{0.67}$Sr$_{0.33}$MnO$_3$ and La$_{0.67}$Ca$_{0.33}$MnO$_3$ films.
\cite{rajeswari1996} On the other hand, $\gamma$ for the present LSMO film grown on LaAlO$_3$ is almost
an order of magnitude larger than that observed in LSMO grown on MgO.\cite{raquet1999}
This discrepancy might indicate a higher defect and dislocation density in the LSMO film on LaAlO$_3$, although the
lattice mismatch between LSMO and MgO is larger than that between LSMO and LaAlO$_3$. This better film quality would,
however, be consistent with other reports on the high quality of manganite films grown on MgO.\cite{ott2000}
The large difference in $\gamma$ of the two samples in the present study indicates additional noise sources
in the A-site disordered LPBMO film which might be related to inherent local structural distortions, defects
and dislocations. In addition to this, its low $T_p$ of 200~K as compared to 375~K for LSMO suggests that there exists
large spin disorder at $T_p$ and below in LPBMO, which might additionally contribute to the $1/f$-noise.

The magnetic field dependence of the magneto-noise might according to Eq.~(\ref{hooge}) arise from the field
dependence of Hooge's constant $\gamma$ or the charge-carrier concentration $n$ and in case of the magnetoresistance,
see Eq.~(\ref{drude}), arise from the field dependence of $n$ or the relaxation time $\tau$.
The agreement between magnetoresistance and magneto-noise
strongly indicates that the carrier density is actually the crucial parameter. This means that the magneto-noise and
magnetoresistance curves should be interpreted as curves showing the change of the inverse carrier concentration with field.
This conclusion is partially supported by evidence from Hall-effect measurements
reporting a field independent Hall mobility\cite{matl1998} and a charge-carrier density minimum in
the vicinity of the Curie temperature.\cite{westerburg2000}
It is likely that these thin  films contain significant levels of quenched disorder such that two competing phases
are present in the samples. The application of a magnetic field leads to the growth of the itinerant phase having a
higher carrier density.

In conclusion, we have observed a direct correspondence between magneto-noise and magnetoresistance effects
in two largely distinct manganite systems, namely, (La$_{0.5}$Pr$_{0.2}$)Ba$_{0.3}$MnO$_3$ and
La$_{0.7}$Sr$_{0.3}$MnO$_3$. Using general arguments on the dependence of the noise and resistivity on the carrier
concentration these observations establish that the $1/f$-electrical noise and the electrical resistivity originate
from charge-carrier density fluctuations. This result is in agreement with both main scenarios discussed in the introduction,
i.e.~charge carrier segregation into polaronic and itinerant phases or the bi-polaron model.
We hope that this method might be further refined for detailed studies of the charge-carrier
density evolution with magnetic field, temperature and doping. An analysis of noise processes within the aforementioned
theoretical models would be desirable for a further understanding.

This work was supported by the DFG under Contract No. DFG ES 86/7-3
within the Forschergruppe ``Oxidi\-sche Grenzfl\"achen''.

\end{document}